\documentclass[12pt]{article}
\pdfoutput=1
\usepackage{amsmath,amssymb}
\usepackage{physics}
\usepackage{bm}
\usepackage[nosort]{cite}
\usepackage{hyperref}

\renewcommand{\title}[1]{\vbox{\center\LARGE{#1}}\vspace{5mm}}
\renewcommand{\author}[1]{\vbox{\center#1}\vspace{5mm}}
\newcommand{\address}[1]{\vbox{\center\em#1}}
\newcommand{\email}[1]{\vbox{\center\tt#1}\vspace{5mm}}

\numberwithin{equation}{section}

\begin{document}

\hypersetup{pageanchor=false}
\begin{titlepage}

\begin{center}

\title{Mass-Flow Invariance of \texorpdfstring{\(Q\)}{Q}-Cohomology\\
in BMN Matrix Quantum Mechanics}
\author{
Chi-Ming Chang$^{a,b}$,
Zhengyuan Du$^{a,c}$,
Sarthak Duary$^a$,\\
Kangning Liu$^{a,c}$ and
Yi-Xiao Tao$^{a,c}$
}

\address{${}^a$Yau Mathematical Sciences Center (YMSC), Tsinghua University, Beijing, China}

\address{${}^b$Beijing Institute of Mathematical Sciences and Applications (BIMSA), Beijing, China}

\address{${}^c$Department of Mathematical Sciences, Tsinghua University, Beijing, China}

\email{cmchang@tsinghua.edu.cn, duzy21@mails.tsinghua.edu.cn\\
sarthakduary@tsinghua.edu.cn, lkn22@mails.tsinghua.edu.cn\\
taoyx21@mails.tsinghua.edu.cn}

\end{center}

\vfill

\begin{abstract}
We study the dependence of the dynamical supercharges of BMN matrix quantum
mechanics on the mass parameter \(\mu\).  Taking the \(\mu\)-derivative at fixed
canonical matrix variables, we show that the sixteen-component supercharge
evolves by the adjoint action of a Hermitian quadratic bosonic operator
\(\mathcal K\), together with the spinor-space factor \(i\gamma^{123}\).
After projection to a \(\gamma^{123}\)-eigenspace, this flow integrates to a
finite similarity transformation.  For the nilpotent component
\(Q(\mu)=\mathcal Q^4_-(\mu)\), one obtains
\(Q(\mu)=M(\mu,\mu_0)Q(\mu_0)M(\mu,\mu_0)^{-1}\), giving an algebraic
mass-flow non-renormalization statement for the \(Q\)-cohomology.  The
corresponding Hilbert-space statement has an analytic qualification, parallel
to Witten's argument for supersymmetric quantum mechanics: \(M\) is non-unitary
and unbounded, so its action on the normalizable domain must be controlled.  We
formulate a small-step criterion by
comparing the quadratic growth of \(M\) with the Gaussian falloff of BMN
oscillator wavefunctions within each component \(\mu>0\) or \(\mu<0\).  As a
concrete check, we evaluate this condition in the \(N=2\) theory, whose two
vacuum sectors are built on the trivial vacuum and the irreducible fuzzy-sphere
vacuum.  We also compute the induced \(Q_{\rm BPS}\)-action on the corresponding
BPS letters: in the trivial sector it agrees with the standard BMN-sector
BPS-letter differential of \(\mathcal N=4\) SYM, while in the irreducible sector
it vanishes.
\end{abstract}

\vfill

\end{titlepage}
\hypersetup{pageanchor=true}

\tableofcontents

\section{Introduction}

BMN matrix quantum mechanics \cite{Berenstein:2002jq} is a massive deformation
of the BFSS matrix model \cite{Banks:1996vh}.  It has
sixteen dynamical supercharges and a discrete set of supersymmetric vacua.  Its
perturbative spectrum and oscillator conventions were studied systematically in
\cite{Dasgupta:2002hx,Kim:2002zg}, and its \(SU(4|2)\) superalgebra and
protected multiplets were analyzed in \cite{Dasgupta:2002ru,Kim:2002if}.

A standard way to isolate protected states is to choose a nilpotent supercharge
\(Q\) and compute the corresponding \(Q\)-cohomology.  In four-dimensional
\(\mathcal N=4\) SYM, this point of view has been used to organize
\(1/16\)-BPS operators \cite{Grant:2008sk,Chang:2013fba}.  More recently,
finite-\(N\) \(Q\)-cohomology has become a practical tool for identifying
non-graviton and black-hole microstate operators, the so-called fortuitous
operators
\cite{Chang:2022mjp,Choi:2022ng,Choi:2023qbm,Choi:2023fnbh,Chang:2024hcf}.
The BMN truncation of the \(Q\)-complex therefore gives a finite-dimensional
testing ground for these questions
\cite{Kim:2003pwm,Choi:2023qbm,Choi:2023fnbh,Gadde:2025bmn,Gaikwad:2025doublegauge},
and it is also closely related to recent finite-\(N\) studies of the BMN
Witten index, including the all-vacuum-sector index computation
\cite{Chang:2024mwi,Chang:2026allsectors}.

These developments make parameter dependence in \(Q\)-cohomology more than a
formal issue.  In \(\mathcal N=4\) SYM, recent work has emphasized that the
classical \(Q\)-cohomology need not agree with the cohomology of the
quantum-corrected, \(g_{\rm YM}\)-dependent supercharge
\cite{Chang:2025sdq,Choi:2025klb,Budzik:2025lcs,Choi:2025frd}.
For the BMN matrix model, the corresponding parameter is the mass \(\mu\), and
the natural question is whether the cohomology of a nilpotent dynamical
supercharge component depends on it.  The parameter \(\mu\) enters the
Hamiltonian through the quadratic mass terms and the Myers term, and it enters
the dynamical supercharges through terms linear in \(\mu\).  A priori, this
raises two related issues:
the explicit \(\mu\)-dependence of the differential \(Q(\mu)\) could change the
algebraic cohomology, and the \(\mu\)-dependent Hamiltonian affects the
Hilbert-space realization, including the normalizability domain on which the
cohomology is defined.

The main observation of this paper is that the \(\mu\)-dependence of the
dynamical supercharges is generated by an inner flow.  More precisely, if
\(\mathcal Q(\mu)\) denotes the spinor-valued dynamical supercharge, then
\begin{equation}
\frac{\partial \mathcal Q}{\partial \mu}
=
i\gamma^{123}[\mathcal K,\mathcal Q],
\qquad
\mathcal K
=
\frac{1}{6R}\Tr(X^iX^i)
-\frac{1}{12R}\Tr(X^aX^a).
\label{eq:intro-main-result}
\end{equation}
The factor \(i\gamma^{123}\) acts only on the suppressed \(SO(9)\) spinor index.
Thus \(\mathcal Q(\mu)\) undergoes a combined spinor-space action and adjoint
flow.  On a fixed \(\gamma^{123}\)-eigenspace this becomes an ordinary
similarity transformation.

We specialize to the nilpotent differential
\begin{equation}
Q(\mu)=\mathcal Q^4_-(\mu),
\end{equation}
which lies in the \(-i\) eigenspace of \(\gamma^{123}\).  Its finite mass-flow
takes the form
\begin{equation}
Q(\mu)
=
M(\mu,\mu_0)Q(\mu_0)M(\mu,\mu_0)^{-1},
\qquad
M(\mu,\mu_0)=e^{(\mu-\mu_0)\mathcal K}.
\label{eq:intro-conjugation}
\end{equation}
At the algebraic level, this identifies the two cohomology groups: if
\(Q(\mu_0)\psi=0\), then \(M\psi\) is \(Q(\mu)\)-closed, and exact states map
to exact states.  In this sense, the result is a non-renormalization statement
for the \(\mu\)-dependence of BMN \(Q\)-cohomology: the mass-dependent
deformation of the differential is generated by an inner flow rather than by a
change of cohomology.

This algebraic isomorphism still has an analytic qualification.  The operator
\(M\) is not unitary; with the Hermitian convention for \(\mathcal K\), it is
an exponential of a real quadratic form in the bosonic coordinates.
Following the logic of Witten's conjugation argument for supersymmetric quantum
mechanics \cite{Witten:1982df,Witten:1982im}, the formal cohomology
isomorphism should be used only on a domain where \(M\) and \(M^{-1}\)
preserve normalizability.  In the BMN model this is plausible for sufficiently
small changes at nonzero \(\mu\), because the Hamiltonian supplies Gaussian
large-field decay in the asymptotic commuting directions.  A finite change of
nonzero \(\mu\) can then be built from small steps.  The point \(\mu=0\) is
exceptional because the quadratic mass terms vanish and the BFSS flat
directions re-emerge.
Within this domain, the standard Hodge-theoretic identification of
Hilbert-space \(Q\)-cohomology with BPS states annihilated by \(Q\) and
\(Q^\dagger\) implies that the BPS states represented by this \(Q\)-cohomology
cannot lift as \(\mu\) varies.

The rest of the paper is organized as follows.  Section~\ref{sec:mass-flow}
fixes conventions for the BMN Hamiltonian and supercharges and derives the
infinitesimal and finite mass-flow generated by \(\mathcal K\).
Section~\ref{sec:q-cohomology} applies the result to \(Q\)-cohomology and
explains the analytic condition inherited from Witten's argument.
Section~\ref{sec:n2-example} treats \(N=2\) as a concrete example, first
checking normalizability of \(M\Omega\) and then computing the induced
\(Q_{\rm BPS}\)-action on BPS letters in the two vacuum sectors: in the trivial
sector it coincides with the standard BMN-sector BPS-letter differential of
\(\mathcal N=4\) SYM, while in the irreducible sector it vanishes.
Section~\ref{sec:discussion} summarizes the main implications.  The appendices
spell out the \(SO(3)\times SO(6)\) spinor decomposition, review the supercharge
anticommutators used to identify a nilpotent differential, record a
dimensionless-coupling form of the supercharge, and finally give the
normal-mode and BPS-projection details behind Section~\ref{sec:n2-example}.

\section{BMN conventions and the mass-flow generated by \texorpdfstring{\(\mathcal K\)}{K}}
\label{sec:mass-flow}
\label{sec:bmn-setup}

\subsection{Hamiltonian and dynamical supercharge}

The Hamiltonian of the BMN matrix quantum mechanics
\cite{Berenstein:2002jq,Dasgupta:2002hx} is
\begin{equation}
\begin{split}
H
&=
R\,\Tr\bigg[
\frac{1}{2}\sum_{I=1}^{9}(\Pi^{I})^{2}
-\frac{1}{4\ell_P^{6}}\sum_{I,J=1}^{9}[X^{I},X^{J}]^{2}
-\frac{1}{2\ell_P^{3}}\Psi^{T}\gamma^{I}[X^{I},\Psi]
\bigg]
\\
&\quad
+\frac{R}{2}\,\Tr\bigg[
\left(\frac{\mu}{3R}\right)^{2}\sum_{i=1}^{3}(X^{i})^{2}
+
\left(\frac{\mu}{6R}\right)^{2}\sum_{a=4}^{9}(X^{a})^{2}
+
i\frac{\mu}{4R}\Psi^{T}\gamma^{123}\Psi
+
i\frac{2\mu}{3R\ell_P^{3}}\epsilon_{ijk}X^{i}X^{j}X^{k}
\bigg].
\end{split}
\label{eq:BMNhamiltonian}
\end{equation}
The nine bosonic matrices are split as
\begin{equation}
i=1,2,3,\qquad a=4,\ldots,9,\qquad I=1,\ldots,9.
\end{equation}
The corresponding spinor-valued dynamical supercharge, in the conventions of
\cite{Dasgupta:2002hx,Kim:2002zg}, is
\begin{equation}
\begin{split}
\mathcal Q
&=
\sqrt{R}\,\Tr\Bigg(
\Pi^a\gamma^a\Psi
+
\Pi^i\gamma^i\Psi
-\frac{i}{2\ell_P^3}[X^i,X^j]\gamma^{ij}\Psi
-\frac{i}{2\ell_P^3}[X^a,X^b]\gamma^{ab}\Psi
\\
&\hspace{2.0cm}
-\frac{i}{\ell_P^3}[X^i,X^a]\gamma^{ia}\Psi
-\frac{\mu}{3R}X^i\gamma^{123}\gamma^i\Psi
+\frac{\mu}{6R}X^a\gamma^{123}\gamma^a\Psi
\Bigg).
\end{split}
\label{eq:BMNsupercharge}
\end{equation}
Here \(\mathcal Q\) denotes the sixteen-component dynamical supercharge.  We
reserve the plain symbol \(Q\) for a chosen nilpotent component, used later as
the cohomological differential.  Appendix~\ref{app:projected-supercharges}
summarizes the \(SO(3)\times SO(6)\) decomposition used to select this
component, while Appendix~\ref{app:dimensionless-coupling} records a
dimensionless parametrization of \eqref{eq:BMNsupercharge} in terms of
\(\lambda=g^{-2/3}=\mu\ell_P^2/R\).

\subsection{Infinitesimal generator from canonical identities}

The bosonic canonical commutation relations are
\begin{equation}
\begin{split}
[X^i_{kl},\Pi^j_{mn}]
&=
i\delta^{ij}\delta_{kn}\delta_{lm},
\\
[X^a_{kl},\Pi^b_{mn}]
&=
i\delta^{ab}\delta_{kn}\delta_{lm}.
\end{split}
\label{eq:canonical}
\end{equation}
The fermionic canonical anticommutation relations will not be needed in this
derivation, because the operator \(\mathcal K\) contains only the bosonic
coordinates.

At fixed canonical variables \(X,\Pi,\Psi\), the only \(\mu\)-dependent terms
in \(\mathcal Q\) are the two linear mass-deformation terms.  Therefore
\begin{equation}
\begin{split}
\frac{\partial \mathcal Q}{\partial\mu}
&=
\sqrt R\,\Tr\left(
-\frac{1}{3R}X^i\gamma^{123}\gamma^i\Psi
+\frac{1}{6R}X^a\gamma^{123}\gamma^a\Psi
\right).
\end{split}
\label{eq:dQdmu}
\end{equation}
We now define the bosonic quadratic operator
\begin{equation}
\mathcal K
=
\frac{1}{6R}\Tr(X^iX^i)
-\frac{1}{12R}\Tr(X^aX^a).
\label{eq:K}
\end{equation}
Evaluating the canonical commutators \eqref{eq:canonical} directly on the
momentum-dependent terms in \eqref{eq:BMNsupercharge}, and noting that the
remaining terms contain only coordinate fields and fermions, we find
\begin{equation}
\begin{split}
[\mathcal K,\mathcal Q]
&=
i\sqrt R\,\Tr\left(
\frac{1}{3R}X^i\gamma^i\Psi
-\frac{1}{6R}X^a\gamma^a\Psi
\right).
\end{split}
\label{eq:KQ}
\end{equation}
Acting on the implicit spinor index with \(i\gamma^{123}\) gives precisely
\eqref{eq:dQdmu}.  Thus
\begin{equation}
\boxed{
\frac{\partial \mathcal Q}{\partial\mu}
=
i\gamma^{123}[\mathcal K,\mathcal Q]
}.
\label{eq:main-result-K}
\end{equation}
This is the basic infinitesimal form of the mass-flow.

\subsection{Finite mass-flow and non-unitarity}

Equation \eqref{eq:main-result-K} is a first-order linear differential
equation in \(\mu\) for the spinor-valued operator \(\mathcal Q\).  It is useful
to write it in terms of the adjoint action
\begin{equation}
\operatorname{ad}_{\mathcal K}(\mathcal O)
=
[\mathcal K,\mathcal O].
\end{equation}
Since \(\mathcal K\) and \(\gamma^{123}\) are independent of \(\mu\), no
path-ordering is needed.  If \(\mathcal Q(\mu_0)\) is the spinor-valued
supercharge at a reference value \(\mu_0\), then
\begin{equation}
\boxed{
\mathcal Q(\mu)
=
\exp\left(
i(\mu-\mu_0)\gamma^{123}\operatorname{ad}_{\mathcal K}
\right)
\mathcal Q(\mu_0)
}.
\label{eq:finite-flow-superoperator}
\end{equation}
Equivalently, this means the series
\begin{equation}
\begin{split}
\mathcal Q(\mu)
&=
\sum_{n=0}^{\infty}
\frac{(\mu-\mu_0)^n}{n!}
\left(i\gamma^{123}\right)^n
\operatorname{ad}_{\mathcal K}^{\,n}
\left(\mathcal Q(\mu_0)\right).
\end{split}
\end{equation}
This is the finite form of \eqref{eq:main-result-K} for the full
unprojected \(SO(9)\) spinor supercharge.

With the Euclidean \(SO(9)\) Clifford convention
\(\{\gamma^I,\gamma^J\}=2\delta^{IJ}\), one has
\begin{equation}
(\gamma^{123})^2=-1.
\end{equation}
After complexifying the spinor space, we may project onto the
\(\gamma^{123}\) eigenspaces,
\begin{equation}
P_{\pm}
=
\frac{1}{2}\left(1\mp i\gamma^{123}\right),
\qquad
\gamma^{123}P_{\pm}
=
\pm i P_{\pm}.
\end{equation}
Here the sign on \(P_\pm\) labels the \(\gamma^{123}\) eigenspace; it is not an
\(SO(3)\) spinor index.
Projecting the differential equation with \(P_\pm\) gives
\begin{equation}
\frac{\partial (P_\pm\mathcal Q)}{\partial\mu}
=
\mp[\mathcal K,P_\pm\mathcal Q].
\end{equation}
This projected equation integrates to an ordinary finite similarity
transformation,
\begin{equation}
\boxed{
P_\pm\mathcal Q(\mu)
=
e^{\mp(\mu-\mu_0)\mathcal K}
P_\pm\mathcal Q(\mu_0)
e^{\pm(\mu-\mu_0)\mathcal K}
}.
\label{eq:finite-flow-projected}
\end{equation}
Thus the finite mass-flow is naturally the spinor-space action of
\(i\gamma^{123}\) combined with the adjoint action of \(\mathcal K\), while
each \(\gamma^{123}\)-eigencomponent is related at different values of
\(\mu\) by a standard similarity transformation.

This similarity should not be mistaken for a unitary equivalence.  Indeed,
with the usual Hermiticity convention \(X^I=(X^I)^\dagger\), the operator
\(\mathcal K\) in \eqref{eq:K} is Hermitian, so
\(e^{\mp\Delta\mu\,\mathcal K}\) is not unitary for real nonzero
\(\Delta\mu=\mu-\mu_0\).  The finite relation
\eqref{eq:finite-flow-projected} should therefore be understood as an
invertible similarity transformation, rather than a unitary equivalence of
Hilbert spaces.

\section{\texorpdfstring{\(Q\)}{Q}-cohomology and Witten's conjugation argument}
\label{sec:q-cohomology}

The finite-flow formula gives a formal conjugation between nilpotent
supercharges at different values of the mass parameter.  The cohomological
question is when this conjugation is an honest map on the relevant Hilbert
space; this is the setting of Witten's conjugation argument
\cite{Witten:1982df,Witten:1982im}.  We choose the nilpotent differential
defining the \(Q\)-cohomology to be the single component
\begin{equation}
Q(\mu)
=
\mathcal Q^4_-(\mu).
\label{eq:chosen-differential}
\end{equation}
With the conventions of Appendix~\ref{app:projected-supercharges},
\(\mathcal Q^A_\alpha\) belongs to the \(-i\) eigenspace of
\(\gamma^{123}\), namely the \(P_-\mathcal Q\) projection.  The same-chirality
anticommutator reviewed in Appendix~\ref{app:nilpotent-supercharges} then
implies directly
\begin{equation}
Q(\mu)^2=0.
\end{equation}
Thus \(Q(\mu)\) is a well-defined differential.
For a nilpotent supercharge, the corresponding BPS Hilbert space is
represented by states obeying \(Q\psi=Q^\dagger\psi=0\), equivalently by
\(\ker\{Q,Q^\dagger\}\) in each fixed charge sector.  Under the usual
Hodge-theoretic identification, this space is isomorphic to the Hilbert-space
\(Q\)-cohomology.  Thus an isomorphism of Hilbert-space \(Q\)-cohomologies
implies that the BPS states represented by this \(Q\)-cohomology cannot lift as
the mass parameter is varied.

For this chosen component, the projected BMN mass-flow gives the formal
conjugating operator
\begin{equation}
M(\mu,\mu_0)
=
e^{(\mu-\mu_0)\mathcal K},
\qquad
Q(\mu)
=
M(\mu,\mu_0)Q(\mu_0)M(\mu,\mu_0)^{-1}.
\label{eq:witten-conjugation-form}
\end{equation}
The operator \(M\) then gives a map between the two cohomology groups.  Indeed,
if \(Q(\mu_0)\psi=0\), then
\begin{equation}
Q(\mu)\big(M\psi\big)
=
MQ(\mu_0)\psi
=
0.
\end{equation}
Moreover, if \(\psi\) is exact at \(\mu_0\), say
\(\psi=Q(\mu_0)\chi\), then
\begin{equation}
M\psi
=
MQ(\mu_0)\chi
=
Q(\mu)M\chi
\end{equation}
is exact at \(\mu\).  The inverse map is \(M^{-1}\).  Thus, whenever
\(M\) is a well-defined invertible operator on the relevant space of states, it
induces an isomorphism
\begin{equation}
H^\bullet(Q(\mu_0))
\cong
H^\bullet(Q(\mu)).
\end{equation}
Therefore, at the purely algebraic level, Witten's conjugation argument gives
an isomorphism of \(Q\)-cohomologies at different nonzero values of \(\mu\).

There is, however, an important analytic qualification.  The situation is
similar to Witten's discussion of the operator appearing in Eq.~(17) of
\cite{Witten:1982df}.  There too the conjugating operator is not unitary, and
the algebraic conjugation formula is not by itself enough.  One must check that
the exponential is an admissible operator: acting on normalizable finite-energy
states, it should again produce normalizable finite-energy states.
Equivalently, the exponential growth from \(M\) must not overwhelm the
large-field decay of the wavefunctions.

In the BMN case, this test has a simple interpretation.  Using
\eqref{eq:witten-conjugation-form} together with \eqref{eq:K}, the
conjugating factor is the coordinate-space multiplier
\[
M(\mu,\mu_0)
=
\exp\left[
(\mu-\mu_0)\left(
\frac{1}{6R}\Tr(X^iX^i)
-\frac{1}{12R}\Tr(X^aX^a)
\right)
\right].
\]
The operator \(M\) is therefore an exponential of a quadratic expression in the
bosonic coordinates.  It is unbounded, and it grows exponentially in some
directions of field space.  This does not automatically invalidate the
conjugation argument, because the BMN Hamiltonian with \(\mu_0\neq0\) already
gives Gaussian decay in the slowest asymptotic directions.  Along the commuting
matrix directions, where the quartic commutator potential and the Myers term do
not improve the asymptotics, the Hamiltonian reduces asymptotically to harmonic
oscillators with frequencies \(|\mu_0|/(3R)\) for \(X^i\) and
\(|\mu_0|/(6R)\) for \(X^a\).  Hence the large-field falloff of low-energy
wavefunctions is of the schematic form
\begin{equation}
\Psi_{\mu_0}
\sim
\exp\left[
-\frac{|\mu_0|}{6R}\Tr(X^iX^i)
-\frac{|\mu_0|}{12R}\Tr(X^aX^a)
\right],
\end{equation}
up to stronger decay in noncommuting directions.

The analytic condition behind the BMN mass-flow argument is therefore that the
quadratic growth introduced by \(M\) must be smaller than the Gaussian decay
already present in the \(\mu_0\)-Hamiltonian.  The schematic comparison gives
the one-sided condition \(|\mu-\mu_0|<|\mu_0|\) for \(M(\mu,\mu_0)\) to act on
the \(\mu_0\)-Hilbert domain.  For a two-sided cohomology isomorphism, the
inverse map \(M(\mu_0,\mu)\) must obey the same condition with the endpoints
interchanged.  Thus the local Hilbert-domain condition is
\begin{equation}
|\mu-\mu_0|<\min(|\mu_0|,|\mu|).
\label{eq:small-step-condition}
\end{equation}
We check this explicitly in the two \(N=2\) perturbative vacuum sectors in
Section~\ref{sec:n2-example}.  A large change in \(\mu\) need not be made in a
single step.  Just as in Witten's example one changes the coupling by repeated
allowed conjugations, here one may connect any two positive values of \(\mu\),
or any two negative values of \(\mu\), by a finite sequence of sufficiently
small steps.

This also explains why \(\mu=0\) is special.  At \(\mu=0\), the quadratic BMN
confinement disappears and one reaches the BFSS asymptotic regime with flat
commuting directions.  The Gaussian falloff used above is then no longer
available, and states can in principle enter from, or escape to, infinity in
field space.  Thus the conjugation argument should be used only along paths in
parameter space that do not cross \(\mu=0\).

\section{Example: \texorpdfstring{\(N=2\)}{N=2}}
\label{sec:n2-example}

This section serves two purposes.  First, we use the \(N=2\) theory as an
explicit check of the Hilbert-domain condition derived in
Section~\ref{sec:q-cohomology}.  Second, independently of this analytic check,
we compute the supercharge induced on the BPS-letter algebra in each vacuum
sector.  This second computation is an additional algebraic output: it records
finite-letter data useful for a direct BMN \(Q\)-cohomology analysis.

For \(N=2\), the BMN vacua are labelled by the two partitions
\cite{Berenstein:2002jq,Dasgupta:2002hx}
\begin{equation}
2=1+1,
\qquad
2=2.
\label{eq:N2-vacuum-partitions}
\end{equation}
Thus there are only two perturbative vacuum sectors: the trivial vacuum
\(X=0\), with unbroken \(U(2)\), and the irreducible fuzzy-sphere vacuum built
from the two-dimensional irreducible representation of \(SU(2)\).  For the
BPS-letter computations below, we write \(P_{\rm BPS}\) for the projection that
keeps the algebra generated by the BPS creation letters, together with the
conjugate annihilation operators needed to evaluate the projected supercharge,
and set \(Q_{\rm BPS}=P_{\rm BPS}Q\).

The ground-state wavefunctions written below should be understood in the
large-\(|\mu_0|\) perturbative limit around the chosen vacuum.  In this limit
the fluctuation Hamiltonian is approximated by its quadratic part, and the
bosonic ground state is a product of Gaussians for the normal modes.  We use
these leading harmonic wavefunctions to test how \(M\) changes the Gaussian
falloff; interaction corrections are not included in the explicit formulas.
The same test applies to finite perturbative excitations: their wavefunctions
are obtained by multiplying the bosonic Gaussian by polynomials in the
normal-mode coordinates, together with finite-dimensional fermion factors.
These polynomial factors do not affect whether the Gaussian integral
converges, so the ground-state estimates below control all finite-letter
perturbative states.

\subsection{The trivial sector}

In the trivial vacuum \(X=0\), the physical bosonic normal modes are the
\(\beta_{1m}\), \(m=-1,0,1\), modes in the \(SO(3)\) directions and the
\(x^a_{00}\), \(a=4,\ldots,9\), modes in the \(SO(6)\) directions.  The
corresponding Gaussian frequencies at mass \(\mu_0\) are
\[
\omega_{\beta,1}=\frac{|\mu_0|}{3R},
\qquad
\omega_{x,0}=\frac{|\mu_0|}{6R}.
\]
In this harmonic approximation, the bosonic Gaussian part of the perturbative
ground state is
\begin{equation}
\Omega^{\rm triv}_{\mu_0}(\beta,x)
=
\mathcal N_{\rm triv}
\exp\left[
-\frac{\omega_{\beta,1}}{2}\sum_{m=-1}^1
\Tr(\beta_{1m}^\dagger\beta_{1m})
-\frac{\omega_{x,0}}{2}\sum_{a=4}^9\Tr\!\left((x^a_{00})^2\right)
\right].
\label{eq:trivial-ground-wavefunction}
\end{equation}
Here the traces are over the \(2\times2\) matrices of the unbroken \(U(2)\).
The constant \(\mathcal N_{\rm triv}\) depends on the normalization of the
bosonic measure, but it will not be needed explicitly: it cancels from the
relative norm below.  The same is true of the overall gauge-group volume when
one imposes the gauge-singlet condition.

The mass-flow operator \(M=e^{\Delta\mu\mathcal K}\), with
\(\Delta\mu=\mu-\mu_0\), changes only the quadratic coefficients of this
Gaussian.  The relative norm is therefore
\begin{equation}
\frac{\norm{M\Omega^{\rm triv}_{\mu_0}}^2}
{\norm{\Omega^{\rm triv}_{\mu_0}}^2}
=
\left(\frac{|\mu_0|}{|\mu_0|-\Delta\mu}\right)^{d_\beta/2}
\left(\frac{|\mu_0|}{|\mu_0|+\Delta\mu}\right)^{d_x/2},
\qquad
d_\beta=12,\quad d_x=24.
\label{eq:trivial-vacuum-M-norm}
\end{equation}
This one-sided Gaussian integral is finite precisely when
\[
|\Delta\mu|<|\mu_0|.
\]
For the inverse map \(M^{-1}=M(\mu_0,\mu)\), the same computation with the
endpoints interchanged gives \(|\Delta\mu|<|\mu|\).  Thus the trivial-sector
calculation reproduces the general two-sided condition
\eqref{eq:small-step-condition}, which is local in the mass parameter.

Having checked the Hilbert-domain condition in this sector, we now record the
second output: the induced differential on the BPS-letter algebra.  With
\(P_{\rm BPS}\) as above, let \(X_{++}\) denote the highest-weight \(SO(3)\)
spinor component of \(X^i\), and let \(\overline{Z}^p\), \(p=1,2,3\), denote the
three complex \(SO(6)\) scalar components.  The bosonic letters \(b,z^p\) and
the fermionic letters \(\eta_p\) are selected by
\begin{equation}
P_{\rm BPS}X_{++}
=\frac{1}{\sqrt{2\omega_{\beta,1}}}\,b,
\qquad
P_{\rm BPS}\overline{Z}^p
=\frac{1}{\sqrt{2\omega_{x,0}}}\,z^p,
\qquad
P_{\rm BPS}\Psi_{+p}=\eta_p,
\qquad
p=1,2,3,
\label{eq:trivial-bps-letters}
\end{equation}
with conjugate annihilators \(b^\dagger,z^{p\dagger},\eta_p^\dagger\).  The
detailed projection is given in Appendix~\ref{app:n2-vacuum-details}; the
resulting \(Q_{\rm BPS}\)-action, after absorbing nonzero normalization
constants into the letters, is
\begin{equation}
\boxed{
Q_{\rm BPS}z^p=0,\qquad
Q_{\rm BPS}\eta_p=\epsilon_{pqr}[z^q,z^r],
\qquad
Q_{\rm BPS}b=[z^p,\eta_p].
}
\label{eq:trivial-bps-Q-action}
\end{equation}
Here repeated \(SU(3)\) flavor indices are summed over the selected three
complex \(SO(6)\) directions.  The nilpotency is already visible in this
intrinsic form.  The only nontrivial check is
\[
Q_{\rm BPS}^2b=[z^p,Q_{\rm BPS}\eta_p]
=
\epsilon_{pqr}[z^p,[z^q,z^r]]=0,
\]
by the Jacobi identity, while \(Q_{\rm BPS}^2z^p=Q_{\rm BPS}^2\eta_p=0\) is
immediate.

This differential is the no-derivative BMN truncation of the \(1/16\)-BPS
\(Q\)-complex in four-dimensional \(\mathcal N=4\) SYM
\cite{Grant:2008sk,Chang:2013fba,Choi:2023qbm,Choi:2023fnbh,Gadde:2025bmn}.
The dictionary is
\[
z^p\leftrightarrow \bar\phi_p,\qquad
\eta_p\leftrightarrow \psi_{p+},\qquad
b\leftrightarrow f_{++},
\]
up to the normalization and sign conventions already absorbed above.  Thus, at
the level of gauge-invariant words built from these BMN letters, the
trivial-vacuum sector has the same \(Q\)-cohomology problem as the BMN-sector
\(Q\)-cohomology of \(\mathcal N=4\) SYM.

\subsection{The irreducible sector}

For the normalizability check in the irreducible sector, we first use
dimensionful fluctuation variables,
\begin{equation}
X^i=X^i_{\rm vac}+Y^i,\qquad
X^i_{\rm vac}=\frac{\mu_0\ell_P^3}{3R}J^i,\qquad
X^a=Y^a,
\label{eq:irreducible-n2-background}
\end{equation}
where \(J^i\) are the dimensionless generators in the two-dimensional irreducible
representation of \(SU(2)\).  The physical bosonic modes are
\(\alpha_{00}\), \(\beta_{jm}\) with \(j=1,2\), and \(x^a_{jm}\) with
\(j=0,1\).  For each spin \(j\), the magnetic quantum number runs over
\(-j\leq m\leq j\).  We use \(\mathcal A^i_{jm}\) and
\(\mathcal B^i_{jm}\) for the trace-one \(\alpha\)- and \(\beta\)-branch
vector harmonics, so \(\alpha_{jm}\) and \(\beta_{jm}\) are the corresponding
mode coefficients.  We use the same complex spherical-basis convention for the
scalar modes.  Thus
\[
\alpha_{jm}^*=(-1)^m\alpha_{j,-m},\qquad
\beta_{jm}^*=(-1)^m\beta_{j,-m},\qquad
(x^a_{jm})^*=(-1)^m x^a_{j,-m},
\]
and the Gaussian quadratic forms below are written as Hermitian norms over the
displayed \(jm\) labels.  Their Gaussian frequencies are
\[
\omega_{\alpha,j}=\frac{|\mu_0|}{3R}(j+1),
\qquad
\omega_{\beta,j}=\frac{|\mu_0|}{3R}j,
\qquad
\omega_{x,j}=\frac{|\mu_0|}{3R}\left(j+\frac12\right).
\]
In the same large-\(|\mu_0|\) approximation, the bosonic Gaussian part of the
perturbative ground state is
\begin{equation}
\begin{split}
\Omega^{\rm irr}_{\mu_0}(\alpha,\beta,x)
&=
\mathcal N^{\rm irr}
\exp\bigg[
-\frac12\sum_{\alpha{\rm\ modes}}\omega_{\alpha,j}
\lvert\alpha_{jm}\rvert^2
-\frac12\sum_{\beta{\rm\ modes}}\omega_{\beta,j}
\lvert\beta_{jm}\rvert^2
-\frac12\sum_{x{\rm\ modes}}\omega_{x,j}\lvert x^a_{jm}\rvert^2
\bigg].
\end{split}
\label{eq:irreducible-ground-wavefunction}
\end{equation}
Here \(\mathcal N^{\rm irr}\) is the corresponding normalization constant for
the Gaussian on the physical fluctuation variables
\(\alpha,\beta,x^a\).  As in the trivial sector, its precise value depends on
the normalization convention for the reduced measure and cancels from the
relative norm below.

The only new feature relative to the trivial sector is that \(\mathcal K\) has
both a constant term on the fuzzy-sphere background and a term linear in the radial
breathing mode \(\alpha_{00}\).  Evaluating the resulting Gaussian integral,
with the constants derived in Appendix~\ref{app:n2-vacuum-details}, gives
\begin{equation}
\boxed{
\begin{aligned}
\frac{\norm{M\Omega^{\rm irr}_{\mu_0}}^2}
{\norm{\Omega^{\rm irr}_{\mu_0}}^2}
&=
e^{\Delta\mu\,\mu_0^2\ell_P^6/(18R^3)}
\exp\left[
\frac{\Delta\mu^2\mu_0^2\ell_P^6/(54R^4)}
{\omega_{\alpha,0}-\Delta\mu/(3R)}
\right]
\\
&\quad\times
\prod_{\alpha{\rm\ modes}}
\left(\frac{\omega_{\alpha,j}}
{\omega_{\alpha,j}-\Delta\mu/(3R)}\right)^{1/2}
\prod_{\beta{\rm\ modes}}
\left(\frac{\omega_{\beta,j}}
{\omega_{\beta,j}-\Delta\mu/(3R)}\right)^{1/2}
\\
&\quad\times
\prod_{x{\rm\ modes}}
\left(\frac{\omega_{x,j}}
{\omega_{x,j}+\Delta\mu/(6R)}\right)^{1/2}.
\end{aligned}
}
\label{eq:irreducible-vacuum-M-norm}
\end{equation}
The products run over the \(N=2\) mode ranges displayed above.  The two-sided
small-step condition \eqref{eq:small-step-condition} makes every denominator
positive.
The extra exponential is the finite contribution from completing the square in
the \(\alpha_{00}\) integral, and it does not change the convergence criterion.
Hence \(M\) and \(M^{-1}\) map finite-letter perturbative states in the
irreducible sector to normalizable states for sufficiently small steps.

Having checked normalizability, we now turn to the BPS-letter algebra in the
irreducible sector.  We continue to use dimensionful BMN field normalizations;
only the fixed \(SU(2)\) matrix harmonics are dimensionless.  Let
\(Y_s\), \(\mathcal B_s\), and \(\mathcal E_s\) denote the highest-weight
scalar, vector, and \(\eta\)-spinor matrix harmonics, respectively, and let
\(\mathcal{Y}_{\ell m}\) denote scalar harmonics before the highest-weight
restriction.  With
\(P_{\rm BPS}\) as above, the bosonic letters \(z^p_s,b_s\) and the fermionic
letters \(\eta_{p;s},\chi_{\frac12}\) are selected by
\begin{equation}
\begin{aligned}
P_{\rm BPS}\overline{Z}^p
&=\sum_{s=0}^{1}\kappa^x_s z^p_s\,Y_s,
&
P_{\rm BPS}Y_{++}
&=\sum_{s=1}^{2}\kappa^\beta_s b_s\mathcal B_s,
\\
P_{\rm BPS}\Psi_{+p}
&=\sum_{s=\frac12}^{\frac32}\kappa^\eta_s\eta_{p;s}\mathcal E_s,
&
P_{\rm BPS}\Psi^4_+
&\propto \chi^\dagger_{\frac12}\mathcal{Y}_{1,0},
&
P_{\rm BPS}\Psi^4_-
&\propto \chi^\dagger_{\frac12}\mathcal{Y}_{1,1}.
\end{aligned}
\label{eq:irreducible-bps-letters-main}
\end{equation}
Here \(p=1,2,3\), the sums run in unit steps, and the constants multiplying the
displayed modes are nonzero.  The final line displays the conjugate annihilator
corresponding to the \(\chi_{\frac12}\) creation letter.
The scalar, vector, and \(\eta\)-spinor matrix harmonics associated with these
highest-weight letters are holomorphic polynomials in \(J_+\).  The explicit
projection in Appendix~\ref{app:n2-vacuum-details} first checks that the
free, or quadratic, supercharge \(Q_2\) has no intrinsic action on the
BPS-letter algebra, including the possible \(\chi^\dagger_{\frac12}\)
annihilation term.  Here \(Q_2\) denotes the part of \(Q\) obtained by
expanding around the irreducible vacuum to first order in fluctuations,
equivalently the piece bilinear in oscillator variables whose anticommutator
gives the quadratic Hamiltonian.
For the remaining nonlinear terms, two cancellations occur: the
\(b^\dagger z\eta\) term is proportional to
\(\Tr([\mathcal B_t^\dagger,Y_r]\mathcal E_u)\), while the
\(zz\eta^\dagger\) term reduces to the commutator
\(\Tr([Y_m,Y_n]\mathcal E_t^\dagger)\).  Both vanish by the trace identities in
Appendix~\ref{app:n2-vacuum-details}.  The final \(Q_{\rm BPS}\)-action is
\begin{equation}
\boxed{
Q_{\rm BPS}z^p_s=0,\qquad
Q_{\rm BPS}\eta_{p;s}=0,\qquad
Q_{\rm BPS}b_s=0,\qquad
Q_{\rm BPS}\chi_{\frac12}=0 .
}
\label{eq:irreducible-bps-Q-action}
\end{equation}
Consequently the projected \(Q\)-cohomology in this sector is simply the
algebra of gauge-invariant combinations of the BPS letters in
\eqref{eq:irreducible-bps-letters-main}, modulo the usual finite-\(N\) trace
relations.  There are no additional \(Q\)-exact identifications inside this
projected letter algebra, because the differential itself vanishes.

\subsection{Perturbative sector preservation}

The estimates above also clarify in what sense \(M\) preserves the two \(N=2\)
vacuum sectors.  In perturbation theory one first chooses a classical vacuum
and expands in local fluctuation coordinates.  Since
\(M=e^{\Delta\mu\mathcal K(X)}\) is multiplication by a nonzero function of the
bosonic coordinates, it leaves the chosen local expansion patch and the
fuzzy-sphere representation label unchanged.  Moreover, \(\mathcal K(X)\) is
built from traces, so \(M\) is gauge invariant and preserves the gauge-singlet
subspace.
In the trivial sector, \eqref{eq:trivial-vacuum-M-norm} comes from a Gaussian
centered at \(X=0\) being mapped to another Gaussian centered at \(X=0\), with
modified widths.  In the irreducible sector,
\eqref{eq:irreducible-vacuum-M-norm} shows that the vacuum is mapped to a
displaced and squeezed Gaussian in the same \(\alpha,\beta,x^a\) fluctuation
variables.  Thus \(M\) need not fix the vacuum vector, but it maps the
perturbative Fock space built over that vacuum into itself on the small-step
domain,
\begin{equation}
\begin{aligned}
M&:\mathcal H_{1+1}^{\rm pert}\to\mathcal H_{1+1}^{\rm pert},\\
M&:\mathcal H_{2}^{\rm pert}\to\mathcal H_{2}^{\rm pert}.
\end{aligned}
\label{eq:N2-M-sector-maps}
\end{equation}
Equivalently, if one formally embeds the local Gaussian wave packets from the
two expansions in a common configuration space, their cross-sector overlaps are
not claimed to vanish exactly at finite \(\mu_0\).  In the large-\(|\mu_0|\)
perturbative regime, however, they are exponentially small: the centers of the
trivial and irreducible packets are separated by order \(|\mu_0|\), while the
oscillator widths are of order \(|\mu_0|^{-1/2}\).  For a fixed step inside the
small-step domain, applying \(M\) keeps the modified quadratic forms positive
and displaces the irreducible Gaussian only within the same fuzzy-sphere branch.
Thus cross-overlaps of the trivial and irreducible packets, with \(M\) applied
to either or both factors, are schematically suppressed as
\(e^{-c|\mu_0|^3}\), with \(c>0\) depending on the step and on normalization
conventions.
This is the perturbative sense in which the sectors are preserved: it concerns
the large-\(|\mu_0|\) local wavepacket sectors, rather than an exact
decomposition of the full finite-\(N\) Hilbert space.

The difference between \eqref{eq:trivial-bps-Q-action} and
\eqref{eq:irreducible-bps-Q-action} is therefore not a change in the chosen
component of the BMN supercharge, but a consequence of the algebra on which that
component is projected.  In the trivial sector the BPS letters are adjoint
matrices and their commutators need not vanish; in the irreducible sector the
noncommutativity is carried by fixed matrix harmonics, and the relevant
highest-weight harmonic factors lie in the commutative holomorphic subalgebra
generated by \(J_+\).

\section{Discussion}
\label{sec:discussion}

We have shown that the \(\mu\)-dependence of the BMN dynamical supercharges is
generated by the Hermitian quadratic operator \(\mathcal K\) in
\eqref{eq:K}.  The spinor-valued supercharge evolves under the
superoperator
\(\exp(i(\mu-\mu_0)\gamma^{123}\operatorname{ad}_{\mathcal K})\).
After restriction to a fixed \(\gamma^{123}\)-eigenspace, this flow becomes an
ordinary similarity transformation.

For the nilpotent component \(Q(\mu)=\mathcal Q^4_-(\mu)\), this gives a
formal cohomology isomorphism between the algebraic \(Q\)-complexes at
different values of the mass parameter.  Equivalently, the algebraic result is
a mass-parameter non-renormalization of the finite-\(N\) BMN \(Q\)-cohomology.
At fixed finite matrix size \(N\),
where there is no ultraviolet renormalization issue, the large-field test
analogous to Witten's supports upgrading this formal conjugation to an actual
isomorphism of Hilbert-space \(Q\)-cohomologies for \(\mu\neq0\), at least
within each connected component \(\mu>0\) or \(\mu<0\).  Consequently, within
the Hilbert-space domain where the non-unitary conjugation is admissible, the
BPS states represented by this \(Q\)-cohomology cannot lift between finite
nonzero values of \(\mu\).

A complete analytic proof of the Hilbert-space cohomology isomorphism would
still require a domain theorem for the unbounded operators \(M\), \(M^{-1}\),
and the chosen nilpotent supercharge.
The continuous argument relates \(\mu>0\) to \(\mu>0\) and \(\mu<0\) to
\(\mu<0\); relating the two signs of \(\mu\) would require a separate symmetry
or equivalence.

The \(N=2\) example in Section~\ref{sec:n2-example} gives a concrete
sector-by-sector application of this conclusion.  In that example the mass-flow
preserves the perturbative vacuum sectors in the large-\(|\mu|\) sense, while
the induced BPS-letter differential is sector dependent: it is the usual
commutator differential in the trivial sector and vanishes in the irreducible
sector.  Thus, within each perturbative sector, the cohomology problem reduces
to the corresponding gauge-invariant letter algebra.

\section*{Acknowledgements}

We thank Edward Witten for discussions.  Chi-Ming Chang is supported by the
NSFC Grant No.~12575075. Zhengyuan Du is supported by the NSFC special fund for theoretical physics No. 12447108 and the national key research
and development program of China No. 2020YFA0713000. Sarthak Duary is supported by the Shuimu Tsinghua Scholar Program of Tsinghua University and the Beijing Natural Science Foundation of China Grant No.~IS25035. Kangning Liu is supported by the NSFC special fund for theoretical physics No. 12447108 and the National Key Research and Development Program of China No. 2020YFA0713000. Yi-Xiao Tao is supported by the NSFC Grant No.~124B2094.

\paragraph*{Note added.}
After the main result of this paper was obtained, Sean Colin-Ellerin informed us
that he and Kyriakos Papadodimas had independently arrived at the same result.
We thank them for sharing this with us, and we have agreed with them to
coordinate the submission of the two papers.

\appendix

\section{\texorpdfstring{\(SO(3)\times SO(6)\)}{SO(3) x SO(6)} spinor conventions and projected supercharges}
\label{app:projected-supercharges}

This appendix collects the spinor conventions used to make the projected
supercharge components explicit.
We decompose the \(SO(9)\) gamma matrices as
\begin{equation}
\gamma^i=\sigma^i\otimes \Gamma,
\qquad
\gamma^a=1\otimes \Gamma^a,
\end{equation}
where \(i=1,2,3\), \(a=4,\ldots,9\), \(\sigma^i\) are Pauli matrices, and
\(\Gamma^a\) are six-dimensional gamma matrices.  We use
\begin{equation}
\Gamma=i\Gamma^4\Gamma^5\cdots \Gamma^9,
\qquad
\gamma^{123}=i\otimes \Gamma.
\end{equation}
The \(SO(9)\) spinor decomposes as
\begin{equation}
\mathbf{16}
\longrightarrow
(\mathbf 2,\mathbf 4)\oplus(\mathbf 2,\overline{\mathbf 4})
\end{equation}
under \(SO(3)\times SO(6)\simeq SU(2)\times SU(4)\).  We write the two
pieces with indices
\begin{equation}
\mathcal Q^A_\alpha,
\qquad
\mathcal Q_{\alpha A},
\qquad
\alpha=\pm,
\qquad
A=1,\ldots,4.
\end{equation}
The \(SO(3)\) spinor index is written in the weight basis, with the convention
\[
\mathcal Q^A_-\equiv \mathcal Q^A_1,
\qquad
\mathcal Q^A_+\equiv \mathcal Q^A_2,
\qquad
\mathcal Q_{-A}\equiv \mathcal Q_{1A},
\qquad
\mathcal Q_{+A}\equiv \mathcal Q_{2A}.
\]
This spinor-weight sign is independent of the \(\gamma^{123}\)-eigenspace sign
appearing on the projectors \(P_\pm\) below.
The \(SO(6)\) chirality matrix is taken to be
\begin{equation}
\Gamma
=
\begin{pmatrix}
-\delta^A_B & 0\\
0 & \delta_A^B
\end{pmatrix}.
\end{equation}
Consequently \(\mathcal Q^A_\alpha\) has \(\gamma^{123}\)-eigenvalue \(-i\),
while \(\mathcal Q_{\alpha A}\) has \(\gamma^{123}\)-eigenvalue \(+i\).  With
the projectors
\begin{equation}
P_\pm=\frac{1}{2}(1\mp i\gamma^{123}),
\qquad
\gamma^{123}P_\pm=\pm iP_\pm,
\end{equation}
this means
\begin{equation}
P_-\mathcal Q
\longleftrightarrow
\mathcal Q^A_\alpha,
\qquad
P_+\mathcal Q
\longleftrightarrow
\mathcal Q_{\alpha A}.
\end{equation}

The \(SO(6)\) vector is represented as an antisymmetric \(SU(4)\) tensor,
\begin{equation}
X_{AB}=\Gamma^a_{AB}X^a,
\qquad
X^{AB}=\Gamma^{aAB}X^a,
\end{equation}
and similarly
\begin{equation}
\Pi_{AB}=\Gamma^a_{AB}\Pi^a,
\qquad
\Pi^{AB}=\Gamma^{aAB}\Pi^a.
\end{equation}
With these conventions, the \(-i\)-eigenspace component of the dimensionful BMN
supercharge \eqref{eq:BMNsupercharge} is
\begin{equation}
\begin{split}
\mathcal Q^A_\alpha(\mu)
=
\sqrt R\,\Tr\bigg(&
\Pi^{AB}\Psi_{\alpha B}
+\Pi^i(\sigma^i)_\alpha{}^\beta\Psi^A_\beta
+\frac{1}{\ell_P^3}X^iX^j\epsilon^{ijk}
(\sigma^k)_\alpha{}^\beta\Psi^A_\beta
\\
&
-\frac{i}{2\ell_P^3}[X^{AB},X_{BC}]\Psi^C_\alpha
-\frac{i}{\ell_P^3}[X^i,X^{AB}]
(\sigma^i)_\alpha{}^\beta\Psi^B_\beta
\\
&
-\frac{i\mu}{3R}X^i(\sigma^i)_\alpha{}^\beta\Psi^A_\beta
+\frac{i\mu}{6R}X^{AB}\Psi_{\alpha B}
\bigg).
\end{split}
\label{eq:Qminus-explicit}
\end{equation}
The \(+i\)-eigenspace component is
\begin{equation}
\begin{split}
\mathcal Q_{\alpha A}(\mu)
=
\sqrt R\,\Tr\bigg(&
\Pi_{AB}\Psi^B_\alpha
-\Pi^i(\sigma^i)_\alpha{}^\beta\Psi_{\beta A}
+\frac{1}{\ell_P^3}X^iX^j\epsilon^{ijk}
(\sigma^k)_\alpha{}^\beta\Psi_{\beta A}
\\
&
-\frac{i}{2\ell_P^3}[X_{AB},X^{BC}]\Psi_{\alpha C}
+\frac{i}{\ell_P^3}[X^i,X_{AB}]
(\sigma^i)_\alpha{}^\beta\Psi^B_\beta
\\
&
-\frac{i\mu}{3R}X^i(\sigma^i)_\alpha{}^\beta\Psi_{\beta A}
-\frac{i\mu}{6R}X_{AB}\Psi^B_\alpha
\bigg).
\end{split}
\label{eq:Qplus-explicit}
\end{equation}

\section{Supercharge anticommutators and nilpotent components}
\label{app:nilpotent-supercharges}

The nilpotency of \(Q(\mu)\) follows from the BMN superalgebra
\cite{Dasgupta:2002ru,Kim:2002if}.  In the original \(SO(9)\) notation, the
anticommutator of the sixteen dynamical supercharges is
\begin{equation}
\{\mathcal Q,\mathcal Q\}
=
2H
-\frac{\mu}{3}\gamma^{123}\gamma^{ij}M^{ij}
+\frac{\mu}{6}\gamma^{123}\gamma^{ab}M^{ab},
\label{eq:SO9-superalgebra}
\end{equation}
where spinor indices are suppressed and the right-hand side is understood as a
matrix acting on the \(SO(9)\) spinor space.  Equivalently, in the
\(SO(3)\times SO(6)\simeq SU(2)\times SU(4)\) notation used in
Appendix~\ref{app:projected-supercharges}, the complex supercharges satisfy
\begin{equation}
\{\mathcal Q^A_\alpha,\mathcal Q^B_\beta\}=0,
\qquad
\{\mathcal Q_{\alpha A},\mathcal Q_{\beta B}\}=0.
\label{eq:same-chirality-anticommutators}
\end{equation}
To write the nonzero mixed anticommutator, we use the chiral \(SO(6)\) spinor
generator in our \(\Gamma\)-matrix notation.\footnote{This footnote only fixes
notation relative to \cite{Dasgupta:2002hx}: their \(g^a\) and
\(g^{\dagger a}\) correspond to our \(\Gamma^a_{AB}\) and
\(\Gamma^{aAB}\), respectively, and their \(g^{ab}\) is the corresponding
chiral \(SO(6)\) generator.}
\begin{equation}
(\Gamma^{ab})^A{}_B
:=
\frac{1}{2}
\left(
\Gamma^{aAC}\Gamma^b_{CB}
-
\Gamma^{bAC}\Gamma^a_{CB}
\right).
\label{eq:chiral-Gamma-ab}
\end{equation}
This is the \(SU(4)\) spinor generator of \(SO(6)\), namely the appropriate
chiral block of the six-dimensional matrix
\(\Gamma^{ab}=\frac{1}{2}[\Gamma^a,\Gamma^b]\).

With this convention, the mixed anticommutator is
\begin{equation}
\{\mathcal Q^A_\alpha,\mathcal Q_{\beta B}\}
=
2\delta^A_B\delta^\alpha_\beta H
-\frac{\mu}{3}\epsilon^{ijk}(\sigma^k)_\beta{}^\alpha\delta^A_B M^{ij}
-\frac{i\mu}{6}\delta^\alpha_\beta(\Gamma^{ab})^A{}_B M^{ab}.
\label{eq:mixed-anticommutator}
\end{equation}
Here \(M^{ij}\) and \(M^{ab}\) generate the \(SO(3)\) and \(SO(6)\) rotations,
respectively.  The index placements follow the conventions of
Appendix~\ref{app:projected-supercharges}; the structural point is that the
Hamiltonian and rotation generators appear only in the mixed anticommutator.

For the differential used in the main text,
\begin{equation}
Q(\mu)=\mathcal Q^4_-(\mu),
\end{equation}
nilpotency follows immediately:
\begin{equation}
Q(\mu)^2
=
\frac{1}{2}\{\mathcal Q^4_-(\mu),\mathcal Q^4_-(\mu)\}
=0.
\end{equation}
Thus \(Q(\mu)\) lies entirely in one \(\gamma^{123}\)-eigenspace.
A mixed choice involving both \(\mathcal Q^A_\alpha\) and
\(\mathcal Q_{\alpha A}\) would instead involve the nonzero mixed anticommutator
\eqref{eq:mixed-anticommutator}.

\section{Dimensionless parametrization of the supercharge}
\label{app:dimensionless-coupling}

For reference, we record a useful dimensionless parametrization of the BMN
supercharge.  For definiteness we take \(\mu>0\); for negative \(\mu\) one may
define the positive coupling using \(|\mu|\) and keep the sign of \(\mu\) in
the mass-deformation terms.  In this appendix a subscript ``phys'' denotes the
dimensionful BMN variables; we also write the supercharge in
\eqref{eq:BMNsupercharge} as \(\mathcal Q_{\rm phys}\).  Unadorned variables
are dimensionless.  We measure time in units of \(R\),
\begin{equation}
t_{\rm phys}=R\,t,
\label{eq:dimensionless-time-R}
\end{equation}
and introduce
\begin{equation}
g^2=\frac{R^3}{\mu^3\ell_P^6},
\qquad
\lambda=g^{-2/3}=\frac{\mu\ell_P^2}{R}.
\label{eq:dimensionless-g-definition}
\end{equation}
Here \(g\) is dimensionless because \(R\), \(\ell_P\), and \(\mu^{-1}\) all
have dimensions of length.

We absorb a common normalization of the rescaled supercharge into the fermion
variable and use
\begin{equation}
X_{\rm phys}^I=\ell_P X^I,
\qquad
\Psi_{\rm phys}
=
\frac{\ell_P}{R}\,\Psi,
\qquad
\Pi_{\rm phys}^I=\frac{1}{\ell_P}\Pi^I.
\label{eq:dimensionless-rescaling-fields}
\end{equation}
The bosonic canonical commutator is preserved:
\begin{equation}
[X^I,\Pi^J]=i\delta^{IJ},
\end{equation}
while the fermion bracket is the one induced by the rescaling in
\eqref{eq:dimensionless-rescaling-fields}.

Starting from the BMN supercharge \(\mathcal Q_{\rm phys}\) in
\eqref{eq:BMNsupercharge}, we rescale the supercharge by \(\sqrt R\), as
appropriate for the time variable in \eqref{eq:dimensionless-time-R}, and then
apply \eqref{eq:dimensionless-rescaling-fields}.  This gives the one-parameter
family
\begin{equation}
\begin{split}
\mathcal Q(\lambda)
=
\Tr\bigg(
&
\Pi^a\gamma^a\Psi
+
\Pi^i\gamma^i\Psi
-\frac{i}{2}[X^i,X^j]\gamma^{ij}\Psi
-\frac{i}{2}[X^a,X^b]\gamma^{ab}\Psi
\\
&\hspace{1.5cm}
-i[X^i,X^a]\gamma^{ia}\Psi
-\frac{\lambda}{3}X^i\gamma^{123}\gamma^i\Psi
+\frac{\lambda}{6}X^a\gamma^{123}\gamma^a\Psi
\bigg).
\end{split}
\label{eq:dimensionless-supercharge-lambda}
\end{equation}
Thus the kinetic and commutator terms are \(\lambda\)-independent, while the
mass-deformation terms are linear in \(\lambda=g^{-2/3}\).  The nilpotent
differential used in the main text is the component
\begin{equation}
Q(\lambda)=\mathcal Q^4_-(\lambda).
\end{equation}

\section{Details for the \texorpdfstring{\(N=2\)}{N=2} example}
\label{app:n2-vacuum-details}

This appendix supplies the normal-mode and BPS-projection details used in
Section~\ref{sec:n2-example}.  The norm computations use the same leading
large-\(|\mu_0|\) harmonic ground-state wavefunctions as in the main text,
with the physical BMN variables and \(\mathcal K\) normalized as in
\eqref{eq:K}.  The BPS-letter projection also uses these dimensionful
oscillator variables; the matrix harmonics are the standard dimensionless
\(SU(2)\) harmonics used in the perturbative BMN spectrum
\cite[Sec.~5]{Dasgupta:2002hx}.  We write \(\mathcal{Y}_{\ell m}\) for the
scalar matrix harmonics.
Throughout this appendix, the \(SO(6)\) complex components are
\[
\overline{Z}^p:=X^{p4},\qquad
Z_p:=X_{p4},\qquad
\Pi^p:=\Pi^{p4},\qquad
\Pi_p:=\Pi_{p4},
\qquad p=1,2,3,
\]
using the \(SU(4)\)-antisymmetric notation of
Appendix~\ref{app:projected-supercharges}.  For \(SO(3)\) vectors we use
\(V_{\alpha\beta}=V^i(\sigma^i)_{\alpha\beta}\), with normalized symmetric
Pauli matrices in the \(\alpha,\beta=\pm\) weight basis; \(V_{++}\) and
\(V_{--}\) are the highest- and lowest-weight components.  We use this notation
for \(X^i\), \(\Pi^i\), and the irreducible-sector fluctuation \(Y^i\).

\subsection{The trivial sector}

Let \(e_m^i\), \(m=-1,0,1\), be the usual orthonormal spherical basis for the
spin-one representation of \(SO(3)\).  In the trivial vacuum the \(SO(3)\) and
\(SO(6)\) coordinates expand as
\begin{equation}
X^i=\sum_{m=-1}^{1}e_m^i\,\beta_{1m},
\qquad
X^a=x^a_{00},
\label{eq:app-trivial-mode-expansion}
\end{equation}
with \(\beta_{1m}^\dagger=(-1)^m\beta_{1,-m}\) and
\((x^a_{00})^\dagger=x^a_{00}\).  The trace normalization is
\begin{equation}
\Tr(X^iX^i)=\sum_{m=-1}^{1}\Tr(\beta_{1m}^\dagger\beta_{1m}),
\qquad
\Tr(X^aX^a)=\sum_{a=4}^{9}\Tr\!\left((x^a_{00})^2\right).
\label{eq:app-trivial-trace-normalization}
\end{equation}
Substituting these modes into the exponent \(\mathcal K\) shows that
\(M=e^{\Delta\mu\mathcal K}\) only shifts the two oscillator widths,
\[
c_\beta=\omega_{\beta,1}-\frac{\Delta\mu}{3R},
\qquad
c_x=\omega_{x,0}+\frac{\Delta\mu}{6R}.
\]
Taking the ratio of the resulting centered Gaussian integrals over
\(3N^2\) real \(\beta\)-components and \(6N^2\) real \(x\)-components gives
\eqref{eq:trivial-vacuum-M-norm}.

The BPS \(SO(3)\) creation letter \(b\), together with its conjugate
annihilator \(b^\dagger\), is defined by the projected components
\begin{equation}
\begin{aligned}
P_{\rm BPS}X_{++}
&=\frac{1}{\sqrt{2\omega_{\beta,1}}}\,b,
&
P_{\rm BPS}X_{--}
&=-\frac{1}{\sqrt{2\omega_{\beta,1}}}\,b^\dagger,
\\
P_{\rm BPS}\Pi_{++}
&=i\sqrt{\frac{\omega_{\beta,1}}{2}}\,b,
&
P_{\rm BPS}\Pi_{--}
&=i\sqrt{\frac{\omega_{\beta,1}}{2}}\,b^\dagger .
\end{aligned}
\label{eq:app-trivial-BPS-XPi}
\end{equation}
For the \(SO(6)\) fields, the BPS creation letters \(z^p\), together with their
conjugate annihilators \(z^{p\dagger}\), are defined by
\begin{equation}
\begin{aligned}
P_{\rm BPS}\overline{Z}^p
&=\frac{1}{\sqrt{2\omega_{x,0}}}\,z^p,
&
P_{\rm BPS}Z_p
&=\frac{1}{\sqrt{2\omega_{x,0}}}\,z^{p\dagger},
\\
P_{\rm BPS}\Pi^p
&=i\sqrt{\frac{\omega_{x,0}}{2}}\,z^p,
&
P_{\rm BPS}\Pi_p
&=-i\sqrt{\frac{\omega_{x,0}}{2}}\,z^{p\dagger}.
\end{aligned}
\label{eq:app-trivial-BPS-ZPi}
\end{equation}
The BPS fermion is
\[
P_{\rm BPS}\Psi_{+p}=\eta_p,
\qquad
P_{\rm BPS}\Psi^p_-=\eta_p^\dagger .
\]
Taking the component \(Q=\mathcal Q^4_-\), all
terms either contain non-BPS modes or annihilate the projected algebra except
for the two terms displayed below.  On the projected letter algebra,
\begin{equation}
Q_{\rm BPS}
=-\frac{i\sqrt R}{2\ell_P^3\omega_{x,0}}\epsilon_{pqr}\Tr(z^pz^q\eta_r^\dagger)
-\frac{i\sqrt R}{2\ell_P^3\sqrt{\omega_{\beta,1}\omega_{x,0}}}
\Tr\!\left([b^\dagger,z^p]\eta_p\right).
\label{eq:app-trivial-QBPS}
\end{equation}
The coefficients in \eqref{eq:app-trivial-QBPS} are nonzero for
\(\mu_0\neq0\); the displayed \(\sqrt R/\ell_P^3\) comes from the
dimensionful BMN supercharge.
Taking graded commutators with \(z^p\), \(\eta_p\), and \(b\), and then
absorbing these constants into the letter normalizations, gives
\eqref{eq:trivial-bps-Q-action}.

\subsection{The irreducible sector}

Around the irreducible fuzzy-sphere vacuum, write the dimensionful fields as
\[
X^i=\frac{\mu_0\ell_P^3}{3R}J^i+Y^i,
\qquad
X^a=Y^a.
\]
After fixing the pure-gauge vector branch to zero, expand directly in the two
trace-one vector-harmonic branches \(\mathcal A^i_{jm}\), \(\mathcal B^i_{jm}\)
and the scalar harmonics \(\mathcal{Y}_{jm}\):
\[
Y^i
=
\sum_{\alpha{\rm\ modes}}\alpha_{jm}\mathcal A^i_{jm}
+
\sum_{\beta{\rm\ modes}}\beta_{jm}\mathcal B^i_{jm},
\qquad
Y^a=\sum_{x{\rm\ modes}}x^a_{jm}\mathcal{Y}_{jm}.
\]
For \(N=2\), the labels are \(\alpha_{00}\), \(\beta_{jm}\) with \(j=1,2\),
and \(x^a_{jm}\) with \(j=0,1\), with \(-j\leq m\leq j\).  These are the
complex spherical-basis coefficients obeying the Hermiticity conditions stated
in the main text.  Orthogonality of the physical vector branches removes
off-diagonal terms, while the background couples only to the radial breathing
mode \(\alpha_{00}\).  The mass-flow exponent is therefore
\begin{equation}
\mathcal K
=
C_{\rm vac}+\ell_\alpha\alpha_{00}
+\frac{1}{6R}\sum_{\alpha{\rm\ modes}}\lvert\alpha_{jm}\rvert^2
+\frac{1}{6R}\sum_{\beta{\rm\ modes}}\lvert\beta_{jm}\rvert^2
-\frac{1}{12R}\sum_{x{\rm\ modes}}\lvert x^a_{jm}\rvert^2 .
\label{eq:app-K-irreducible}
\end{equation}
The first term is the value of \(\mathcal K\) on the fuzzy-sphere background,
while the second comes from the linear coupling to the radial breathing mode.
Using
\(\sum_i\Tr(J^iJ^i)=N(N^2-1)/4\), and then setting \(N=2\), gives
\begin{equation}
C_{\rm vac}=\frac{\mu_0^2\ell_P^6}{36R^3},
\qquad
\ell_\alpha=\frac{\mu_0\ell_P^3\sqrt6}{18R^2}.
\label{eq:app-Cvac-ell-N2}
\end{equation}
Thus \(M=e^{\Delta\mu\mathcal K}\) shifts the \(\alpha\)-, \(\beta\)-, and
\(x\)-mode widths by \(-\Delta\mu/(3R)\), \(-\Delta\mu/(3R)\), and
\(+\Delta\mu/(6R)\), respectively, and adds the linear source
\(\Delta\mu\,\ell_\alpha\alpha_{00}\).  Completing the square in the
\(\alpha_{00}\) integral and integrating the remaining centered Gaussians gives
\eqref{eq:irreducible-vacuum-M-norm}.

For the BPS-letter computation, \(J^i\) are the dimensionless \(SU(2)\)
generators, and \(Y_s:=\mathcal{Y}_{s,s}\), \(\mathcal B_s\), and
\(\mathcal E_s\) denote the relevant matrix harmonics; the dimensionful factors
are carried by the oscillator normalizations such as \(\kappa^x_r\) and
\(\kappa^\beta_s\).  Let
\(J_+=J^1+iJ^2\).  Up to nonzero normalization constants, the highest-weight
scalar, vector, and \(\eta\)-spinor matrix harmonics are
\[
Y_s\propto J_+^s,
\qquad
\mathcal B_s\propto J_+^{s-1},
\qquad
\mathcal E_s\propto J_+^{s-\frac12}.
\]
Here \(\mathcal B_s\) is the matrix part of the highest-weight \(Y_{++}\)
component of the same \(\beta\)-branch vector harmonic \(\mathcal B^i_{s,s}\)
whose coefficients are the \(\beta\)-branch modes above.
They are all holomorphic in \(J_+\).  Hence the holomorphic products commute:
\begin{equation}
[Y_r,Y_s]=0,
\qquad
[Y_r,\mathcal B_s]=0,
\qquad
[Y_r,\mathcal E_s]=0.
\label{eq:app-holomorphic-harmonics}
\end{equation}
The trace identities needed below are
\begin{equation}
\Tr\!\left([\mathcal B_t^\dagger,Y_r]\mathcal E_u\right)=0,
\qquad
\Tr\!\left([Y_r,Y_s]\mathcal E_t^\dagger\right)=0 .
\label{eq:app-irred-trace-identities}
\end{equation}
With these conventions, the BPS projections of the bosonic fields are
\begin{equation}
\begin{aligned}
P_{\rm BPS}\overline{Z}^p
&=\sum_{r=0}^{1}\kappa^x_r\,z^p_r\,Y_r,
&
P_{\rm BPS}Z_p
&=\sum_{r=0}^{1}\kappa^x_r\,z^{p\dagger}_r\,Y_r^\dagger,
\\
P_{\rm BPS}Y_{++}
&=\sum_{s=1}^{2}\kappa^\beta_s\,b_s\,\mathcal B_s,
&
P_{\rm BPS}Y_{--}
&=-\sum_{s=1}^{2}\kappa^\beta_s\,b_s^\dagger\,\mathcal B_s^\dagger,
\end{aligned}
\label{eq:app-irred-boson-projections}
\end{equation}
with
\[
\kappa^x_r=\frac{1}{\sqrt{2N\omega_{x,r}}},
\qquad
\kappa^\beta_s=\frac{1}{\sqrt{2N\omega_{\beta,s}}}.
\]
For the fermions,
\begin{equation}
P_{\rm BPS}\Psi_{+p}
=\sum_{s=\frac12}^{\frac32}\kappa^\eta_s\,\eta_{p;s}\,\mathcal E_s,
\qquad
P_{\rm BPS}\Psi^p_-
=\sum_{s=\frac12}^{\frac32}\kappa^\eta_s\,
\eta^\dagger_{p;s}\,\mathcal E_s^\dagger,
\qquad
\kappa^\eta_s=\frac{1}{\sqrt N}.
\label{eq:app-irred-fermion-projections}
\end{equation}
For \(N=2\) the only \(\chi\)-letter is \(\chi_{\frac12}\).  The corresponding
conjugate annihilation pieces selected by \(P_{\rm BPS}\) are
\begin{equation}
P_{\rm BPS}\Psi^4_+
=
-\frac{1}{\sqrt{3N}}\chi^\dagger_{\frac12}\mathcal{Y}_{1,0},
\qquad
P_{\rm BPS}\Psi^4_-
=
\sqrt{\frac{2}{3N}}\chi^\dagger_{\frac12}\mathcal{Y}_{1,1}.
\label{eq:app-irred-chi-projections}
\end{equation}
With our letter convention, \(\chi^\dagger_{\frac12}\) denotes the conjugate
annihilator of the creation letter \(\chi_{\frac12}\).
Expanding the dimensionful component supercharge around
\(X^i_{\rm vac}=\mu_0\ell_P^3J^i/(3R)\), the pure-background
\(\Psi^4_\pm\) terms cancel between the \(1/\ell_P^3\) commutator pieces and
the \(\mu_0/R\) mass pieces; in components
\[
X^{\rm vac}_{--}=\frac{\mu_0\ell_P^3}{3R}\frac{J_+}{\sqrt2},
\qquad
X^{\rm vac}_{++}=-\frac{\mu_0\ell_P^3}{3R}\frac{J_-}{\sqrt2},
\qquad
X^{\rm vac}_{+-}=\frac{\mu_0\ell_P^3}{3R}\frac{J_3}{\sqrt2}.
\]
The scalar-\(\eta\) terms vanish because the scalar BPS harmonics are
holomorphic in \(J_+\).  The remaining possible \(\chi^\dagger_{\frac12}\)
terms vanish by the endpoint identities
\begin{equation}
\begin{aligned}
\Tr\!\left([J_+,\mathcal B_s]\mathcal{Y}_{1,1}\right)&=0,
\\
\Tr\!\left(\mathcal B_s^\dagger\mathcal{Y}_{1,0}\right)
&=
\Tr\!\left([J_3,\mathcal B_s^\dagger]\mathcal{Y}_{1,0}\right)=0,
\\
\Tr\!\left([\mathcal B_s^\dagger,J_-]\mathcal{Y}_{1,1}\right)&=0,\qquad s=1,2.
\end{aligned}
\label{eq:app-irred-Q2-chi-identities}
\end{equation}
Thus the full free, or quadratic, contribution obeys
\begin{equation}
P_{\rm BPS}Q_2=0.
\label{eq:app-irred-Q2-projection-zero}
\end{equation}
Substituting these projections into the nonlinear terms, the possible
\(b^\dagger z\eta\) and \(zz\eta^\dagger\) contributions are removed by the two
trace identities in \eqref{eq:app-irred-trace-identities}, respectively.  Hence
\begin{equation}
Q_{\rm BPS}=0 .
\label{eq:app-irred-QBPS-before-zero}
\end{equation}
This gives the zero differential displayed in
\eqref{eq:irreducible-bps-Q-action}.


\begin{thebibliography}{99}
\sloppy
\hbadness=2000

\bibitem{Berenstein:2002jq}
D.~Berenstein, J.~M.~Maldacena, and H.~Nastase,
``Strings in flat space and pp waves from \(\mathcal N=4\) super Yang Mills,''
JHEP \textbf{0204} (2002) 013,
arXiv:hep-th/0202021.

\bibitem{Banks:1996vh}
T.~Banks, W.~Fischler, S.~H.~Shenker, and L.~Susskind,
``M theory as a matrix model: A conjecture,''
Phys. Rev. D \textbf{55} (1997) 5112--5128,
arXiv:hep-th/9610043.

\bibitem{Dasgupta:2002hx}
K.~Dasgupta, M.~M.~Sheikh-Jabbari, and M.~Van Raamsdonk,
``Matrix perturbation theory for M-theory on a pp-wave,''
JHEP \textbf{0205} (2002) 056,
arXiv:hep-th/0205185.

\bibitem{Kim:2002zg}
N.~Kim and J.~Plefka,
``On the spectrum of pp-wave matrix theory,''
Nucl. Phys. B \textbf{643} (2002) 31--48,
arXiv:hep-th/0207034.

\bibitem{Dasgupta:2002ru}
K.~Dasgupta, M.~M.~Sheikh-Jabbari, and M.~Van Raamsdonk,
``Protected multiplets of M-theory on a plane wave,''
JHEP \textbf{0209} (2002) 021,
arXiv:hep-th/0207050.

\bibitem{Kim:2002if}
N.~Kim and J.-H.~Park,
``Superalgebra for M-theory on a pp-wave,''
Phys. Rev. D \textbf{66} (2002) 106007,
arXiv:hep-th/0207061.

\bibitem{Kim:2003pwm}
N.~Kim, T.~Klose, and J.~Plefka,
``Plane-wave matrix theory from \(\mathcal N=4\) super Yang-Mills on \(R\times S^3\),''
Nucl. Phys. B \textbf{671} (2003) 359--382,
arXiv:hep-th/0306054.

\bibitem{Grant:2008sk}
L.~Grant, P.~A.~Grassi, S.~Kim, and S.~Minwalla,
``Comments on \(1/16\) BPS quantum states and classical configurations,''
JHEP \textbf{0805} (2008) 049,
arXiv:0803.4183.

\bibitem{Chang:2013fba}
C.-M.~Chang and X.~Yin,
``\(1/16\) BPS states in \(\mathcal N=4\) SYM,''
Phys. Rev. D \textbf{88} (2013) 106005,
arXiv:1305.6314.

\bibitem{Chang:2022mjp}
C.-M.~Chang and Y.-H.~Lin,
``Words to describe a black hole,''
JHEP \textbf{02} (2023) 109,
arXiv:2209.06728.

\bibitem{Choi:2022ng}
S.~Choi, S.~Kim, E.~Lee, and J.~Park,
``The shape of non-graviton operators for \(SU(2)\),''
JHEP \textbf{09} (2024) 029,
arXiv:2209.12696.

\bibitem{Choi:2023qbm}
S.~Choi, S.~Kim, E.~Lee, S.~Lee, and J.~Park,
``Towards quantum black hole microstates,''
JHEP \textbf{11} (2023) 175,
arXiv:2304.10155.

\bibitem{Choi:2023fnbh}
J.~Choi, S.~Choi, S.~Kim, J.~Lee, and S.~Lee,
``Finite \(N\) black hole cohomologies,''
JHEP \textbf{12} (2024) 029,
arXiv:2312.16443.

\bibitem{Chang:2024hcf}
C.-M.~Chang and Y.-H.~Lin,
``Holographic covering and the fortuity of black holes,''
arXiv:2402.10129.

\bibitem{Gadde:2025bmn}
A.~Gadde, E.~Lee, R.~Raj, and S.~Tomar,
``Probing non-graviton spectra in \(\mathcal N=4\) SYM via BMN truncation and S-duality,''
JHEP \textbf{02} (2026) 026,
arXiv:2506.13887.

\bibitem{Gaikwad:2025doublegauge}
A.~Gaikwad, T.~Kibe, S.~van Leuven, and K.~Mathieson,
``To gauge or to double gauge? Matrix models, global symmetry, and black hole cohomologies,''
JHEP \textbf{05} (2026) 133,
arXiv:2512.02103.

\bibitem{Chang:2024mwi}
C.-M.~Chang,
``Witten index of BMN matrix quantum mechanics,''
SciPost Phys. \textbf{19} (2025) 147,
arXiv:2404.18442.

\bibitem{Chang:2026allsectors}
C.-M.~Chang, S.~Duary, and K.~Liu,
``Finite-\(N\) BMN index across all vacuum sectors,''
arXiv:2605.25560.

\bibitem{Chang:2025sdq}
C.-M.~Chang and Y.-H.~Lin,
``Violation of \(S\)-duality in classical \(Q\)-cohomology,''
arXiv:2510.24008.

\bibitem{Choi:2025klb}
J.~Choi and E.~Lee,
``Konishi lifts a black hole,''
arXiv:2511.09519.

\bibitem{Budzik:2025lcs}
K.~Budzik and J.~Kulp,
``Loop corrected supercharges from holomorphic anomalies,''
arXiv:2512.07771.

\bibitem{Choi:2025frd}
J.~Choi and S.~Kim,
``Fortuity and relevant deformation,''
arXiv:2512.12674.

\bibitem{Witten:1982df}
E.~Witten,
``Constraints on supersymmetry breaking,''
Nucl. Phys. B \textbf{202} (1982) 253--316.

\bibitem{Witten:1982im}
E.~Witten,
``Supersymmetry and Morse theory,''
J. Differential Geom. \textbf{17} (1982) 661--692.

\end{thebibliography}
\end{document}